\documentclass[useAMS, usenatbib]{mn2e}
\usepackage{amssymb,amsmath}
\usepackage{graphicx}
\usepackage{xcolor}
\usepackage{multirow}

\title[Non-kinematic double-peaks]
{SDSS J083253.18+064316.7: one strange object with double-peaked narrow 
H$\alpha$ but single-peaked narrow H$\beta$}
\author[Zhang X.-G.]
       {Xue-Guang Zhang$^{1,2}$\thanks{xgzhang@pmo.ac.cn}\\
       $^1$Purple Mountain Observatory, Chinese Academy of Sciences,
             2 Beijing XiLu, NanJing, JiangSu, 210008, P. R. China \\
       $^2$Chinese Center for Antarctic Astronomy, NanJing, 
             JiangSu, 210008, P. R. China}

\date{}
\def\LaTeX{L\kern-.36em\raise.3ex\hbox{a}\kern-.15em
    T\kern-.1667em\lower.7ex\hbox{E}\kern-.125emX}

\begin{document}
\pagerange{\pageref{firstpage}--\pageref{lastpage}} \pubyear{}
\maketitle
\label{firstpage}

\begin{abstract}
    In this letter, we firstly report one unique object SDSS J0832+0643 with 
particular features of narrow balmer emission lines: double-peaked narrow 
H$\alpha$ but single-peaked narrow H$\beta$. The particular features can not be 
expected by currently proposed kinematic models for double-peaked narrow emission 
lines, because the proposed kinematic models lead to similar line profiles of 
narrow balmer emission lines. However, due to radiative transfer effects, the 
non-kinematic model can be naturally applied to well explain the particular 
features of narrow balmer emission lines: larger optical depth in H$\alpha$ than 
10 leads to observed double-peaked narrow H$\alpha$, but smaller optical depth 
in H$\beta$ around 2 leads to observed single-peaked narrow H$\beta$. Therefore, 
SDSS J0832+0643 can be used as strong evidence to support the non-kinematic 
model for double-peaked narrow emission lines.
\end{abstract}

\begin{keywords}
Galaxies:Active -- Galaxies:nuclei -- Galaxies:Seyfert -- quasars:Emission lines
\end{keywords}

\section{Introduction}

   Dual supermassive black holes could be one inevitable stage, if co-evolution 
of supermassive black holes and host galaxies was accepted (Silk \& Rees 1998, 
Di Matteo, Springel \& Hernquist 2005, Hopkins et al. 2006). Once separations 
of central two supermassive black holes are around kilo-pcs scale, double-peaked 
narrow emission lines could be expected, if there are significant velocities 
along the line-of-sight of the dual supermassive black holes. Therefore, 
double-peaked narrow emission lines can be treated as one indicator of candidate 
dual supermassive black holes with separations about kilo-pcs, such as the 
following reported well studied candidates among AGNs with double-peaked narrow 
emission lines, SDSS J1048 (Zhou et al. 2004) DEEP2 J1420+5259 (Gerke et 
al. 2007), EGSD2 J1415 (Comerford et al. 2009a), COSMOS J1000 (Comerford et 
al. 2009, Blecha et al. 2013, Wrobel, Comerford \& Middelberg 2014), 
SDSS J1316 (Xu \& Komossa 2009), SDSS J1517 (Rosario et al. 2010), SDSS J1715 
(Comerford et al. 2011), SDSS J1502 (Fu et al. 2011b), SDSS J0952 (Mcgurk 
et al. 2011), SDSS J1426 (Barrows et al. 2012), SDSS J1502 (Fu et al. 2012), 
3C316 (An et al. 2013), NDWFS J1432 and J1433 (Comerford et al. 2013), 
SDSS J1108, J1146, J1131 and J1332 (Liu et al. 2013), SDSS J1323 (Woo et 
al. 2014), etc.. And then, some samples of objects with double-peaked 
narrow emission lines are reported as candidates for objects with dual 
supermassive black holes, such as Wang et al. (2009), Smith et al. (2010) 
and Ge et al. (2012) etc..
  
    Besides the dual supermassive black holes, there are another kinematic 
models which can be applied to explain observed double-peaked narrow emission 
lines, such as radial outflows and rotating disk systems (Liu et al. 2010, 
Fischer et al. 2011, Shen et al. 2011, Comerford et al. 2012, Fischer et al. 
2013, etc.). Through high-quality NIR images discussed in Fu et al. (2011) 
for 50 double-peaked narrow line AGNs, in Shen et al. (2011) for 31 double-peaked 
narrow line AGNs and in Fu et al. (2012) for 42 double-peaked narrow line AGNs, 
one conclusion has been reported that among their small sample of double-peaked 
narrow line AGNs, scenarios involving a single AGN can produce double-peaked 
narrow emission lines with considerations of gas kinematics for large part of 
the double-peaked narrow line AGNs, only a small part of double-peaked narrow 
line AGNs probably contain dual supermassive black holes.  
    
     Furthermore, besides the proposed kinematic models for double-peaked 
narrow emission lines, there is one another non-kinematic model, the model 
based on radiative transfer effects in optically thick sources (more recent 
detailed discussions on the non-kinematic model can be found in Elitzur, 
Ramos \& Ceccarelli 2012), which can also be well applied to explain observed 
double-peaked narrow emission lines. And moreover, as what have been discussed 
in literature, it is generally impossible to disentangle the effects of 
kinematics and line opacities in observed double-peaked emission lines.

     However, we should note that there are some different model expected 
features of double-peaked narrow emission lines, based on proposed kinematic 
and non-kinematic models. On the one hand, the kinematical models should lead 
to the similar line profiles of double-peaked narrow balmer emission lines, 
because the kinematic models have the same effects on narrow balmer emission 
lines. On the other hand, due to much different line opacities for narrow 
balmer emission lines, the non-kinematic model could lead to some different 
double-peaked narrow balmer emission lines. However, in the literature, there 
are so far no reliable results reported on non-kinematic origin of 
double-peaked narrow emission lines. Here, in this letter, we firstly report 
such a unique object with double-peaked narrow H$\alpha$ but single-peaked 
narrow H$\beta$, which will provide strong evidence to support the non-kinematic 
model leading to the different line profiles of narrow balmer emission lines. 
In section 2, we give our main results and discussions on the object SDSS 
J0832+0643. And in section 3, we give our final conclusions.

\section{Main Results}

\begin{figure}
\centering\includegraphics[width = 8cm,height=4cm]{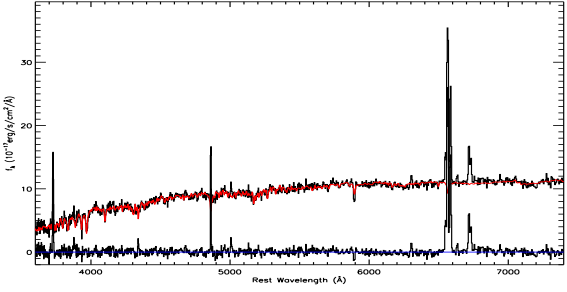}
\caption{SDSS spectrum and line spectrum of SDSS J0832+0643. From top to 
bottom, solid line in black shows the SDSS spectrum, solid line in red shows 
the determined stellar contributions, solid line in black shows the pure line 
spectrum, and solid line in blue shows $f_\lambda=0$. Here, in order to show 
more clearer plots, the observed spectrum and the line spectrum are smoothed 
by average values per 4 data points.
}
\label{spec}
\end{figure}

   In order to check properties of broad balmer emission lines of double-peaked 
narrow line AGNs, emission lines have been carefully checked for all the 
double-peaked narrow line objects reported in Ge et al. (2012), which will 
provide further information on origin of double-peaked narrow emission lines 
(Zhang 2015, submitted to MNRAS). When we check shifted velocities of broad 
balmer lines relative to double-peaked narrow balmer lines, we find SDSS 
J083253.18+064316.7 (=SDSS J0832+0643) with the single-peaked narrow H$\beta$. 
Except the SDSS J0832+0643, no other objects can be found with so strange narrow 
balmer lines. Then, by the following fitted results for narrow balmer emission 
lines and the following Kolmogorov-Smirnov statistic results, we can confirm 
that it is true for double-peaked narrow H$\alpha$ but single-peaked narrow 
H$\beta$.

   The observed SDSS spectrum is shown in Figure~\ref{spec} for SDSS J0832+0643. 
In order to show more clearer properties of narrow balmer emission lines, the 
stellar lights in the spectrum should be firstly subtracted. Here, the more 
recent proposed pPXF method (Penalized Pixel-Fitting method discussed in 
Cappellar \& Emsellem 2004) with 224 MILES (Medium resolution INT Library of 
Empirical Spectra, Vazdekis et al. 2012) template spectra is applied to 
determine the stellar contributions, because the regularization technique 
included in the pPXF method could lead to more smoother solutions on spectral 
decomposition.  Here, The 224 MILES template spectra have stellar ages from 
0.0631Gyr to 17.7828Gyr and have [M/H] from -2.32 to 0.22. The determined stellar 
contributions by the pPXF method are shown in Figure~\ref{spec}.

\begin{figure*}
\centering\includegraphics[width = 14cm,height=6cm]{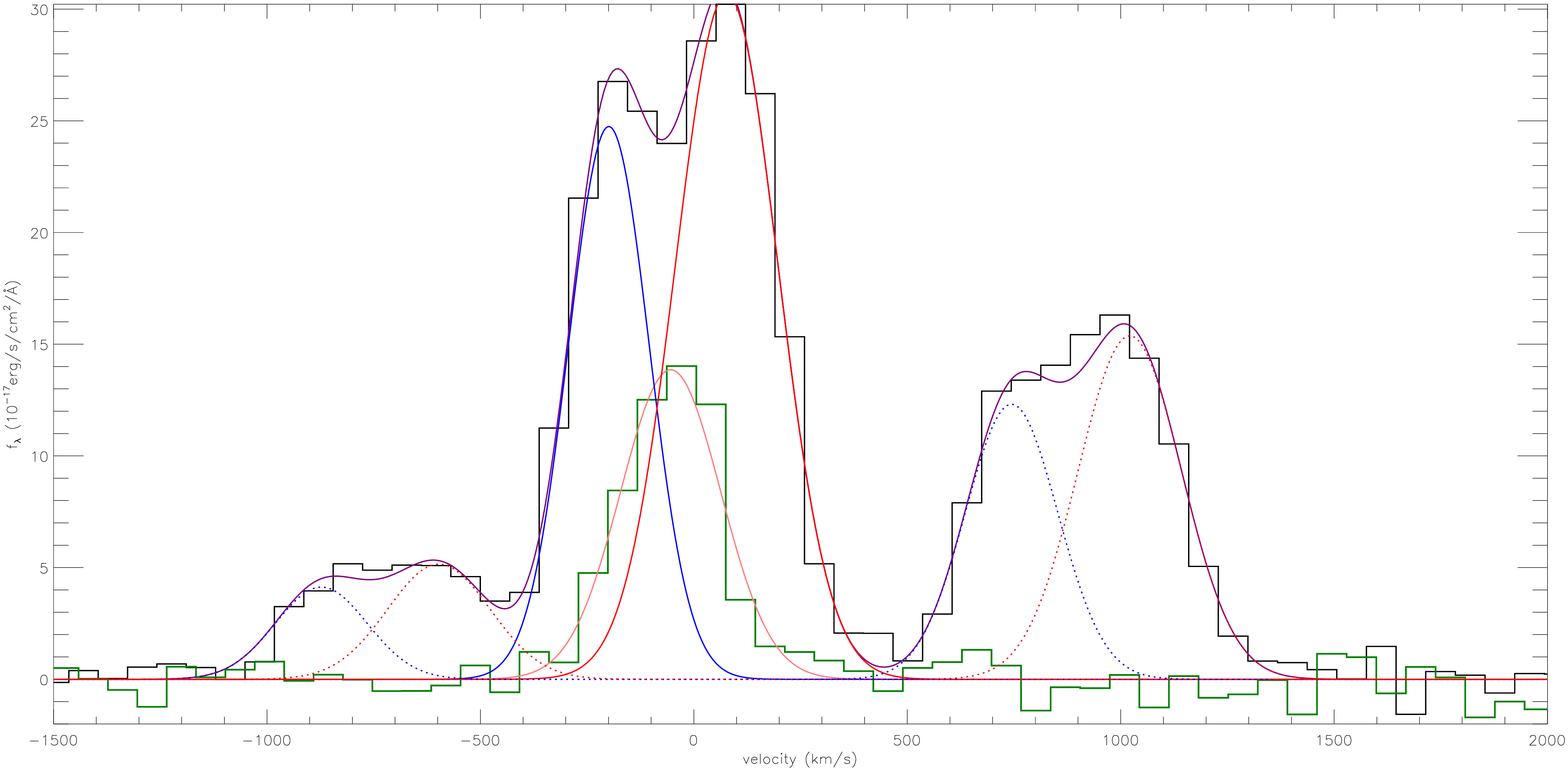}
\caption{properties of narrow balmer emission lines in the velocity space. 
Solid line in black shows the line spectrum (stellar contributions having 
been subtracted) around H$\alpha$ (velocities calculated relative to center 
wavelength of 6564.61\AA). Solid line in purple shows the 
best fitted result by six narrow gaussian functions for the double-peaked 
narrow H$\alpha$ and the double-peaked [N~{\sc ii}] doublet. Solid lines 
in blue and in red show the blue component and the red component of the 
double-peaked narrow H$\alpha$ respectively. Dotted lines in blue and in 
red show the blue components and the red components of the double-peaked 
[N~{\sc ii}] doublet. Solid line in dark green shows the observed single-peaked 
narrow H$\beta$ in the velocity space (velocities calculated relative to 
center wavelength of 4862.68\AA), and thick solid line in pink represents 
the best fitted result for the narrow H$\beta$ by one gaussian 
function.
}
\label{line}
\end{figure*}

   Based on the pPXF determined results, the stellar velocity dispersion of 
host galaxy is about ${\rm 155\pm17 km/s}$ in SDSS J0832+0643, which is 
consistent with the reported value $\rm 158\pm15 km/s$ determined by the 
STARLIGHT code (Cid Fernandes et al. 2005) in Ge et al. (2012). The coincident 
results indicate our determined stellar components are reliable. Then, after 
stellar lights being removed, emission lines can be well checked. And we can 
find that SDSS J0832+0643 is one type 2 objects with pure narrow emission lines: 
strong narrow balmer lines and much weak [O~{\sc iii}]$\lambda5007\AA$ line. 
Figure~\ref{line} shows the fitted results for narrow emission lines by gaussian 
functions. In this paper, we mainly focus on the lines of narrow H$\alpha$, 
narrow H$\beta$ and [N~{\sc ii}]$\lambda6548, 6583\AA$ doublet. The determined 
line parameters are listed in Table 1. Here, there is one important 
point we should note. When our procedure is applied to describe each narrow 
emission line by two gaussian functions, we try to fit the narrow H$\beta$ by 
the similar line profile of the double-peaked narrow H$\alpha$. However, we find 
that FWHM (full width at half maximum) is about 510${\rm km/s}$ for the narrow 
H$\alpha$, but is only 270${\rm km/s}$ for the narrow H$\beta$. The much 
different FWHMs strongly indicate different line profiles of narrow balmer 
lines. So that the narrow H$\beta$ is fitted without considerations of line 
profile of double-peaked narrow H$\alpha$. Furthermore, although two gaussian 
components are applied to fit the narrow H$\beta$ in our procedure, the fitted 
result for narrow H$\beta$ leads to zero flux of one component, which indicates 
one gaussian function is good enough to describe the narrow H$\beta$. Furthermore, 
the two-sided Kolmogorov-Smirnov statistic technique is applied to check 
difference between line profiles of narrow balmer lines. Based on the line 
profile of narrow H$\beta$ from the observed spectrum with stellar contributions 
being subtracted and the line profile of narrow H$\alpha$ created by the two 
gaussian components listed in Table 1, the Kolmogorov-Smirnov statistic gives 
one probability about 0.0015 to support the assumption that narrow H$\alpha$ 
and narrow H$\beta$ have the similar line profile. And moreover, based on the 
line profile of narrow H$\beta$ from the observed spectrum with stellar 
contributions being subtracted and the fitted result for the narrow H$\beta$ 
by the one gaussian components listed in Table 1, the Kolmogorov-Smirnov 
statistic gives one probability about 83\% to support the assumption that the 
narrow H$\beta$ has one gaussian line profile. Therefore, one gaussian function 
rather than two gaussian functions is preferred to describe the narrow H$\beta$.

   Before proceeding further, simple comparisons are shown with reported results 
of double-peaked narrow H$\alpha$ in Ge et al. (2012) for SDSS J0832+0643. In 
Ge et al. (2012), the reported line parameters of the double-peaked narrow 
H$\alpha$ are as follows: $\sigma\sim2.2\pm0.1\AA$ and 
$flux\sim145.8\pm10.5\times10^{-17}{\rm erg/s/cm^2}$ for the blue component, 
and $\sigma\sim2.6\pm0.1\AA$ and $flux\sim210.4\pm11.9\times10^{-17}
{\rm erg/s/cm^2}$ for the red component respectively. It is clear that the 
results on line width and line flux listed in Table 1 are well consistent with 
reported results in Ge et al. (2012). And moreover, the peak separation is 
about 5.99$\pm$0.26\AA in Ge et al. (2012), which is also well consistent with 
our result 6$\pm$0.26\AA. The results indicate our determined line parameters 
of double-peaked narrow H$\alpha$ are reliable. Furthermore, 
there is one point we should note. In Ge et al. (2012), when emission lines 
were determined, there was one strong criterion that each narrow emission 
line had the similar double-peaked line profile, which led to mathematic 
determined line parameters of two fake components of narrow 
[O~{\sc iii}]$\lambda5007\AA$ and narrow H$\beta$ of SDSS J0832+0643, and 
then led to classification of one two Type 2 AGNs of SDSS J0832+0643. 
However, it is clear that there are much different line profiles of narrow 
balmer lines, which indicate the criterion applied in Ge et al. (2012) is not 
valid for the narrow balmer lines of SDSS J0832++0643. So that, in this 
paper, we do not show further discussions on classification of SDSS 
J0832++0643. Based on the results discussed above, we can find that the 
narrow H$\alpha$ is double-peaked but the narrow H$\beta$ is single-peaked 
in SDSS J0832+0643. This is so far the only one reported unique object with 
the double-peaked narrow H$\alpha$ but the single-peaked narrow H$\beta$. 
And moreover, the single-peaked narrow H$\beta$ is not corresponding to blue 
or red component of the double-peaked narrow H$\alpha$, through their 
determined center wavelengths listed in Table 1. Then, we can check the 
theoretical models for the particular features of narrow balmer emission lines.

\begin{table}
%\small
%\renewcommand{\tabcolsep}{1.1mm}
%\begin{minipage}{17.5cm}
\caption{Line parameters}
\begin{tabular}{llccc}
\hline
line & mark & $\lambda_0$ & $\sigma$ & flux \\
\hline
\multirow{2}{*}{H$\alpha$} & blue & 6560.3$\pm$0.13 & 2.1$\pm$0.1 & 128$\pm$8 \\
      & red &  6566.3$\pm$0.13 & 2.6$\pm$0.1 & 201$\pm$9 \\
\hline
\multirow{2}{*}{[N~{\sc ii}]} & blue & 6580.9$\pm$0.13 & 2.4$\pm$0.1 & 73$\pm$5 \\
       & red &  6586.9$\pm$0.13 & 2.7$\pm$0.1 & 102$\pm$5 \\
\hline
H$\beta$ & & 4861.8$\pm$0.2 & 1.9$\pm$0.2 & 66$\pm$9 \\
\hline
\end{tabular}\\
Notice:
The first column shows which line is measured, the second column shows 
which component of the line is measured. The third column shows the center 
wavelength in unit of \AA\ for the component, the fourth column shows 
the second moment in unit of \AA\ for the component and the fifth column 
shows the flux in unit of ${\rm 10^{-17}erg/s/cm^2}$ of the component. 
\end{table}

   Under the kinematic models for double-peaked narrow emission lines, 
there is no way to expect so different line profiles of narrow balmer emission 
lines. Even, with considerations of uncertainties, we could assume that 
the single-peaked narrow H$\beta$ could be related to the red component of 
double-peaked narrow H$\alpha$. It looks that the dual supermassive 
black holes could be applied to explain the observed features of narrow balmer 
lines with disappearance of the blue component of the narrow H$\beta$, if there 
was much larger dust extinction for the blue components of the intrinsic 
double-peaked narrow balmer emission lines. Under the assumed case with large 
dust extinction for the blue components, we would expect the $E(B-V)$ could 
be not less than 1 (flux ratio of H$\alpha$ to H$\beta$ larger than 10). 
However, for the red components of double-peaked narrow balmer lines, the flux 
ratio of H$\alpha$ to H$\beta$ is only 3.05, one common standard value. If the 
dust extinction in the blue component was due to abundant dusts due to galaxy 
mergering, it is natural to expect that there should be similar effects of 
mergering on the two components. If we assume that the single-peaked narrow 
H$\beta$ could be related to the blue component of double-peaked narrow 
H$\alpha$, similar results can be found: flux ratio of the red components of 
narrow H$\alpha$ to narrow H$\beta$ is larger than 10, but the flux ratio of 
the blue component is only 1.9. Therefore, the dual supermassive 
black holes could not be one natural choice to explain the observed features 
of double-peaked narrow H$\alpha$ but single-peaked narrow H$\beta$ of SDSS 
J0832+0643.

\begin{figure}
\centering\includegraphics[width = 8cm,height=6cm]{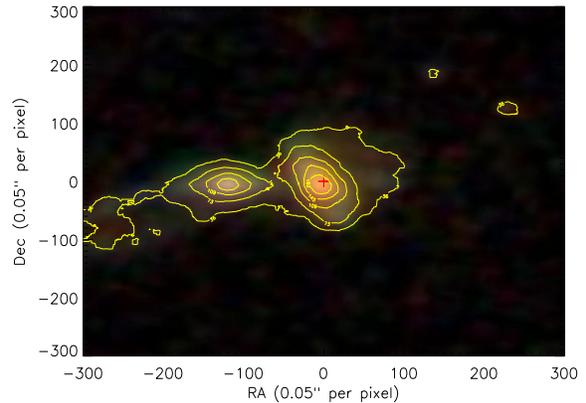}
\caption{SDSS image of SDSS J0832+0643. Red plus shows the
position with RA of 08:32:53.18 and DEC of +06:43:16.7, and each pixel
in the image has a width of 0.05 arcseconds.
}
\label{image}
\end{figure}

    The non-kinematic model can be one natural and reasonable model to 
explain double-peaked narrow H$\alpha$ but single-peaked narrow H$\beta$. 
Based on the measured line flux of total H$\alpha$ to H$\beta$, the flux ratio 
is about 5, which indicate the optical depths in H$\alpha$ and in H$\beta$ are 
about $\tau_{H\alpha}\sim10-11$ and $\tau_{H\beta}\sim0.2\times\tau_{H\alpha}\sim2$ 
respectively, with the following assumptions of optical depth of Ly$\alpha$ 
about $10^{5-6}$ and electron density $N_e\sim10^8 {\rm cm^{-3}}$ as discussed 
in Luna \& Costa (2005). Here, we do not have enough information to give one 
reliable estimation on electron density. But $N_e\sim10^8cm^{-3}$ could be 
accepted, such as the listed lower limit values of electron densities for 
southern symbiotic stars in Luna \& Costa (2005). And moreover, 
the very weak [O~{\sc iii}]$\lambda5007\AA$ line support the high electron 
density to some extent. Furthermore, Figure~\ref{image} shows the SDSS image 
of SDSS J0832+0643: one clear mergering system. Therefore, there could be 
abundant dusts in central region of SDSS J0832+0643, which can naturally 
explain the large flux ratio of narrow H$\alpha$ to narrow H$\beta$.

   Once, the optical depths in H$\alpha$ and in H$\beta$ are simply 
determined, we can give some discussions on different line profiles of narrow 
balmer emission lines. As more recent discussions in Elitzur, Ramos \& 
Ceccarelli (2012), optical depth around 10 can be treated as one boundary 
value: optical depth larger than 10 would lead to double-peaked narrow lines, 
but optical depth around 1 should not lead to double-peaked narrow lines. 
In SDSS J0832+0643, $\tau_{H\alpha}>10$ but $\tau_{H\beta}\sim2$  could lead 
to the double-peaked narrow H$\alpha$ but the single-peaked narrow H$\beta$. 
Furthermore, under the non-kinematic model, the peak separation (about 
270${\rm km/s}$) of the two peaks of narrow H$\alpha$ can be described as 
$\sim2\times \nu_D\times\sqrt{\ln(\tau_{H\alpha})}$, where $\nu_D$ 
represents the broadening velocity due to the thermal motions. Therefore, 
$\nu_D\sim80-90{\rm km/s}$ is enough to create the double-peaked narrow 
H$\alpha$ without contributions from rotating component, which indicates 
the electron temperature is about ${\rm 10^5-10^6K}$. The expected 
$\nu_D$ leads to the expected peak separation of intrinsic double-peaked 
narrow H$\beta$ is only 120${\rm km/s}$, therefore, it is hard to find 
apparent double-peaked features of narrow H$\beta$ because of smaller 
peak separation. It is clear that the non-kinematic model with considerations 
of radiative transfer effects can be applied to well explain the observed 
double-peaked narrow H$\alpha$ but the single-peaked narrow H$\beta$ in 
SDSS J0832+0643.

    Before the end of the section, there are three points we should note. 
First and foremost, the non-kinematic model expected double-peaked narrow 
emission line could be symmetric to some extent. However, the observed 
double-peaked narrow H$\alpha$ of SDSS J0832+0643 is asymmetric to some extent, 
the red component is more stronger than the blue component. The asymmetric 
double-peaked narrow H$\alpha$ indicates another weak component included in 
the narrow emission lines, one probable radial flow component. If the radial 
components were removed, the flux ratio of H$\alpha$ to H$\beta$ should be 
a bit larger than 5, leading to a bit larger $\tau_{H\alpha}$ leading to more 
consistent results with the observed features of narrow balmer lines. 

   Besides, it is hard to find the non-kinematic model expected double-peaked 
narrow emission lines, because much lower electron density in common NLRs 
leading to flux ratio of narrow H$\alpha$ to narrow H$\beta$ around 3  and 
smaller electron temperate leading to much smaller peak separations of expected 
double-peaked narrow emission lines.

\begin{figure}
\centering\includegraphics[width = 8cm,height=8cm]{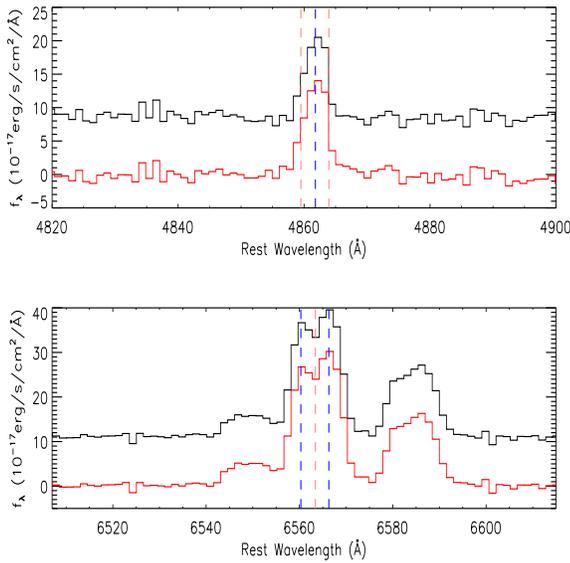}
\caption{Properties of narrow H$\alpha$ (bottom panel) and narrow H$\beta$ 
(top panel). In each panel, solid line in black shows the spectrum before  
subtractions of stellar contributions, and solid line in red shows the 
spectrum after subtractions of stellar contributions. The vertical dotted blue 
lines represent peak positions of the narrow balmer lines, 4861.8\AA\ for the 
single-peaked narrow H$\beta$, and 6560.3\AA\ and 6566.3\AA\ for the 
double-peaked narrow H$\alpha$. In top panel, the vertical pink dashed  
lines show expected peak positions ([6560.3\AA,  6566.3\AA]
$\times4862.68/6564.61$) based on the two peaks of the narrow H$\alpha$. In bottom 
panel, the vertical pink dashed line shows expected peak position 
(4861.8\AA$\times6564.61/4862.68$) based on the peak of the narrow H$\beta$.
}
\label{line2}
\end{figure}

    Last but not least, we give further discussions to confirm that the 
narrow H$\alpha$ and the narrow H$\beta$ have much different line profiles. 
Figure~\ref{line2} shows the line profiles of narrow balmer lines, before and 
after subtractions of stellar contributions. It is clear that even in the SDSS 
spectrum without subtractions of stellar contributions, there are double-peaked 
narrow H$\alpha$, but single-peaked narrow H$\beta$. And moreover, we have 
shown that the stellar velocity dispersion is about 155${\rm km/s}$ (about 
2.5\AA\ around H$\beta$), which is larger than the measured line width of 
narrow H$\beta$. Therefore, there are few effects of subtractions of stellar 
contributions on the single-peaked narrow H$\beta$, and the conclusion is 
reliable that the double-peaked narrow H$\alpha$ and the single-peaked narrow 
H$\beta$ are not fake.

\section{Conclusions}

    Finally, we give our conclusions as follows. First and foremost, SDSS 
J0832+0643 is so far the firstly reported object with particular features of 
narrow balmer emission lines: double-peaked narrow H$\alpha$ but single-peaked 
narrow H$\beta$. Besides, the kinematic models can not explain the observed 
features of narrow balmer emission lines of SDSS J0832+0643. Last but not least, 
the non-kinematic model with considerations of radiative transfer effects can 
be applied to well explain the observed features of narrow balmer emission 
lines of SDSS J0832+0643.

\section*{Acknowledgments}
 Zhang X. G. gratefully acknowledge the anonymous referee for 
giving us constructive comments and suggestions to greatly improve our paper. 
Zhang acknowledges the kind support from the Chinese grant NSFC-11003043
and NSFC-11178003. This paper has made use of the data from the SDSS projects. 
SDSS-IV is managed by the Astrophysical Research Consortium for the Participating 
Institutions of the SDSS Collaboration including the Carnegie Institution 
for Science, Carnegie Mellon University, the Chilean Participation Group, 
Harvard-Smithsonian Center for Astrophysics, Instituto de Astrofisica de 
Canarias, The Johns Hopkins University, Kavli Institute for the Physics 
and Mathematics of the Universe (IPMU) / University of Tokyo, Lawrence 
Berkeley National Laboratory, Leibniz Institut fur Astrophysik Potsdam (AIP),
Max-Planck-Institut fur Astrophysik (MPA Garching), Max-Planck-Institut
fur Extraterrestrische Physik (MPE), Max-Planck-Institut fur Astronomie
(MPIA Heidelberg), National Astronomical Observatory of China, New
Mexico State University, New York University, The Ohio State University,
Pennsylvania State University, Shanghai Astronomical Observatory,
United Kingdom Participation Group, Universidad Nacional Autonoma de
Mexico, University of Arizona, University of Colorado Boulder, University
of Portsmouth, University of Utah, University of Washington, University
of Wisconsin, Vanderbilt University, and Yale University.

\label{lastpage}
\end{document}